\documentclass[twocolumn,showpacs,showkeys,amsmath,amssymb]{revtex4}

\usepackage{graphicx}
\usepackage{dcolumn}
\usepackage{bm}

\newcommand{\be}{\begin{equation}}
\newcommand{\ee}{\end{equation}}
\newcommand{\nn}{\nonumber}
\newcommand{\bea}{\begin{eqnarray}}
\newcommand{\eea}{\end{eqnarray}}
\newcommand{\bfig}{\begin{figure}}
\newcommand{\efig}{\end{figure}}
\newcommand{\bc}{\begin{center}}
\newcommand{\ec}{\end{center}}

\begin{document}
\noindent
PITHA 05/13, IFIC/05-42, \\ ZU-TH 17/05, UCLA/05/TEP/27 \\ \\

\title{QCD Corrections to Static Heavy Quark Form Factors}

\author{W.\ Bernreuther$^a$, R.\ Bonciani$^{b}$, T.\ Gehrmann$^c$, 
R.\ Heinesch$^a$, T.\ Leineweber$^{a,}$\footnote{Now at: Framatome
ANP GmbH,
D-91050 Erlangen, Germany.},
P.\ Mastrolia$^d$, E.\ Remiddi$^e$}
 \affiliation{$^a$ Institut f\"ur Theoretische Physik, RWTH Aachen,
D-52056 Aachen, Germany\\
$^b$  Departament de F\'{\i}sica Te\`orica, IFIC, CSIC -- Universitat de 
Val\`encia, E-46071 Val\`encia,
Spain \\
$^c$ Institut f\"ur Theoretische Physik,
Universit\"at Z\"urich, CH-8057 Z\"urich, Switzerland\\
$^d$ Department of Physics and Astronomy, UCLA,
Los Angeles, CA 90095-1547, U.S.A.\\
$^e$ Dipartimento di Fisica dell'Universit\`a di Bologna and
INFN, Sezione di Bologna, I-40126 Bologna, Italy}

\date{\today}

\begin{abstract}
Interactions of heavy quarks, in particular of top quarks, with
electroweak  gauge bosons are 
expected to be very sensitive to new physics effects related to electroweak
symmetry breaking. These interactions are described by the so-called 
static form factors, which include anomalous magnetic
moments and the effective weak charges. We compute the second-order 
QCD corrections to these static form factors, which turn out to be sizeable 
and need to be taken into account in searches for new 
anomalous coupling effects. 
\end{abstract}

\maketitle



The top quark, so far the heaviest known fundamental particle, 
will serve as an excellent probe of the fundamental
interactions at energies of a few hundred GeV, once more and more
top quark events will be accumulated at the Tevatron and, especially,
after  the CERN Large Hadron Collider (LHC) will  have started
operating.  Even more detailed investigations into top quark 
properties will be possible at a future linear electron-positron collider
(ILC), where top quark pairs can be studied in a very clean and 
well-defined environment.

In view of its large mass the top quark is, in particular, a unique probe
of the dynamics that breaks the electroweak gauge symmetry. If this
mechanism differs from the Higgs mechanism of the standard model (SM) of
particle physics,  observable effects could  be found first in top
quark production and decay. They  may manifest themselves  as deviations of
the top-quark gauge-boson couplings from the values predicted by the
SM (c.f. \cite{Murayama:1996ec,Hill:2002ap} for overviews).

There has been an enormous effort in recent years to investigate 
the potential of top quarks, and also of bottom quarks, to 
new physics effects.  Specifically, the couplings to photons and
$Z$ bosons, which are the subject of this Letter, have been studied
in detail -- both for heavy quark production at hadron colliders
\cite{Baur1,Baur2}
and at a future high-luminosity high-energetic linear electron-positron
collider \cite{Kane:1991bg} (for reviews,
see \cite{tesla,Abe:2001nq}). Indirect constraints, which are rather
tight for $b$ quarks, 
on anomalous contributions to these couplings can be obtained 
also from electroweak precision observables 
measured at LEP \cite{Baur1,Altarelli,eboli,Larios:1999au}.
Many of these studies attempt to use a ``model-independent''
approach in that the effect of new interactions is 
described in terms of effective Lagrangians or, equivalently, in terms
of anomalous couplings of heavy quarks, in particular
to the SM gauge bosons.  In this context the obvious question arises
about the size of these effective couplings within the SM, in
order to assess  the margin of detectability of truly anomalous new
physics effects,  given the experimental sensitivity of some
observable.

In this Letter we investigate the
couplings of heavy quarks, noteably of the $t$ and $b$ quark, to the photon
and $Z$ boson in higher order QCD. In view of the asymptotic freedom
property of the strong interactions and  of the large energy
scale  set by the heavy quark mass, $m_Q \gg \Lambda_{QCD}$, these effective
couplings can be computed perturbatively. 
These couplings do in general depend on the precise kinematics at the 
vertex, which is expressed in terms of 
 the so-called vertex form factors, depending 
on the momentum transfer at the vertex. In many applications, it is justified 
to approximate these form factors by their limits at zero momentum 
transfer, the so-called static form factors. The most prominent static 
form factor is the electromagnetic spin-flipping form factor: the 
anomalous magnetic moment.  
At present, radiative corrections to the static form factors of quarks
are known 
to one loop in the SM \cite{Hollik,Bernabeu}, while pure QED corrections to 
the anomalous magnetic moment are known analytically to three 
loops \cite{LapRem}. 
 As far as these quantities  are concerned, the largest radiative
 corrections are often
those due to QCD. Here we present analytic results for the
static $\gamma Q Q$ and $Z QQ$ form factors to second order in the QCD
coupling $\alpha_s$ and we predict the size of these moments for $t$
and $b$ quarks within the SM.
Moreover we briefly discuss the implications of our results in view of
an existing upper bound on 
the anomalous magnetic moment of the $b$ quark and on future
precision measurements of the $\gamma tt$ and $Z tt$ couplings.

To start with we  define the static form factors of a heavy quark $Q$ by
considering  for definiteness 
the kinematical situation $V^* \to Q(p_1)+ {\bar Q}(p_2)$ ($V^*$ =
off-shell photon or $Z$ boson).   We study the $V^* Q {\bar Q}$ vertex
functions $ \bar{u}(p_1)\Gamma^{\mu,V}_Q(q)v(p_2)$ 
 for on-shell external
quarks in the limit of zero four-momentum
transfer $q=p_1+p_2$. In general this vertex function can be decomposed, using
Lorentz covariance, into six form factors, two of which are
CP-violating. CP violation in  a  flavour-conserving vertex
function is, in the SM,  a tiny, higher-loop induced effect, and we do
not consider it here any further.  
Assuming CP invariance  $\Gamma^{\mu,V}_Q$ 
then depends on  four form factors:
\bea
i\, \Gamma^{\mu,V}_Q  &=&   e\,  \bigg(F_{1,Q}^V \gamma^\mu +
 \frac{i}{2m_Q}F_{2,Q}^V \sigma^{\mu \nu} q_\nu 
\nn \\ && +   G_{1,Q}^V \gamma^\mu \gamma_5 +
  \frac{1}{2m_Q}G_{2,Q}^V \gamma_5 q^\mu \bigg) \, ,\label{decomp}
\eea
where the form factors
 are functions of
$s=q^2$. Further $\sigma^{\mu \nu} = \frac{i}{2}\left[\gamma^\mu,\gamma^\nu
\right]$,  and $e>0$ denotes the positron charge. 
If one considers the matrix element of
the electromagnetic current, the parity-violating part of
Eq. (\ref{decomp}) is to be replaced by $G_Q(\gamma_\mu\gamma_5 s -
2m_Q \gamma_5 q_\mu)$, where $G_Q(0)$ yields the anapole moment of
$Q$.

We recall that for  a reaction involving the $VQQ$
vertices the physical object is the $S$ matrix element, but not, in general, these form
factors. The static quantities $F_{i,Q}^{\gamma}(0)$,
$G_{i,Q}^{\gamma}(0)$ do have a physical meaning: they are defineable as
the residues of the photon  pole in scattering amplitudes in the soft
photon limit, and they are gauge invariant (with respect to the full SM
gauge group) and infrared-finite. Likewise, the $F_{i,Q}^{Z}(m_Z^2)$,
$G_{i,Q}^{Z}(m_Z^2)$, which  determine the $S$ matrix element of the
decay of an on-shell $Z$ boson into a $Q \bar Q$ ($Q = b,c$) quark
pair, are gauge invariant -- but they are not infrared-finite, as the $Q
\bar Q$ state above threshold is degenerate with states containing in
addition soft photons/gluons and/or collinear massless partons. On the
other hand, the static form factors  $F_{i,Q}^{Z}(0)$, $G_{i,Q}^{Z}(0)$
are infrared-finite, but gauge invariant only with respect to QCD or
pure QED.

Here our primary aim is to compute the anomalous magnetic moments of
heavy quarks in QCD, which are physical quantities. In addition we
determine the heavy-quark anomalous weak magnetic moment and its
axial charge to second order in $\alpha_s$. Although these are well-defined objects 
in QCD only, these SM predictions should serve as useful reference
values, in particular for new physics models whose effect on, for
instance, the $e^+e^- \to Q {\bar Q}$ amplitude is essentially confined
to the $ V Q {\bar Q}$ vertices.

To lowest order in the SM couplings, only $F_{1,Q}^{\gamma,Z}$ and 
$G_{1,Q}^{Z}$ are non-zero. The other  form factors are generated by
one-loop
radiative corrections: $G_{1,2,Q}^{\gamma}$ by parity-violating weak
corrections, and $F_{2,Q}^{\gamma,Z}$ and $G_{2,Q}^Z$ by strong and
electroweak corrections. In this Letter we consider the QCD
corrections to these form factors. To lowest order in the electroweak
couplings we have $G_{1,2,Q}^{\gamma}=0$ and  $F_{i,Q}^V = v_Q^V
F_{i,Q}$  $(i=1,2)$, where ${v}_Q^\gamma = Q_Q$ and 
$ {v}_Q^Z =({T_3^Q}/{2}-s_w^2\,Q_Q)/(s_w c_w),$ ${a}_Q = -{T_3^Q}/(2s_w c_w)$
 are the SM vector and axial vector couplings of $Q$
to the $Z$ boson and the photon in units of
$e$, where $s_w(c_w)$ is the sine (cosine) of the weak mixing angle,
$T_3^Q$ the third component of the weak isospin and $Q_Q$ is the charge
of the heavy quark. 

Here we determine these form factors in the static limit to second
order in $\alpha_s$, using the results for non-zero
momentum transfer $s$ given in 
\cite{Bernreuther:2004ih}. 
We work in QCD with $N_l$ massless quarks and one  quark $Q$ with 
mass $m_Q$ (defined in the on-shell scheme). This includes
the case of six quarks with all quarks but the top quark taken to
be massless. The QCD coupling
$\alpha_s = \alpha_s(\mu)$ is defined in the standard $\overline {\rm
  MS}$ scheme in $N_f=N_l+1$ flavor QCD with $\mu$ being the renormalization
scale. 

Because of conservation of the electromagnetic and the neutral vector
current, the QCD corrections to $F_{1,Q}$ vanish for $s\to 0$; i.e.,
 $F_{1,Q}(s=0)=1.$ 
The form factors
$F_{2,Q}$, $G_{1,Q}$, and $G_{2,Q}$ 
become infrared finite in the static limit.  We obtain 
\be
F_{2,Q}(s=0) =
\frac{\alpha_s}{2\pi}\, C_F
+
\left(\frac{\alpha_s}{2\pi}\right)^2
F_{2,Q}^{(2l)} \, ,
\ee
with
\bea
 F_{2,Q}^{(2l)}  =
 \phantom{+}C_F^2\left(
  -\frac{31}{4}+2\,\zeta_2\,(5-6\,\ln(2))+3\,\zeta_3\right)
\nonumber \\  +C_F\,C_A\left( \frac{317}{36}
  +3\,\zeta_2\,(-1+2\,\ln(2))-\frac{3}{2}\,\zeta_3\right)\nonumber
\\
+C_F\,T_F\left(\frac{119}{9}-8\,\zeta_2\right) 
- \frac{25}{9} C_F\,T_F\,N_l + C_F\,\beta_0\,\ln(r_Q),
\eea
where $r_Q={\mu^2}/{m_Q^2}$, $\zeta_n$ is the Riemann zeta function, and
$C_F=(N_c^2-1)/2N_c$, $C_A=N_c$, $T_F=1/2$ with $N_c=3$ being
the number of colors. Further $\beta_0=(11\,C_A-4\,T_F\,(N_l+1))/6$. 
For $G_{1,Q}$ we have, to  order $\alpha_s^2$:
\be
G_{1,Q}^Z= a_Q \left(G_{1,Q}^{(A)}+
  G_{1,Q}^{(B)} \right)\, ,
\ee
where the superscripts $A,B$ denote universal and non-universal
corrections, respectively.
We get for the type A  term: 
\be
G_{1,Q}^{(A)}(s=0) = 1 -
\frac{\alpha_s}{2\pi}\, C_F
+ \left(\frac{\alpha_s}{2\pi}\right)^2G_{1,Q}^{(2l)} \, ,\label{g1_s0_nichtano}
\ee
with
\bea
G_{1,Q}^{ (2l)}  = 
\phantom{+} C_F^2 \left(-\frac{29}{12} +
  8\,\zeta_2\,(1-\ln(2))+2\,\zeta_3\right) \nonumber \\
+ C_F\,C_A \left(
  -\frac{143}{36}+2\,\zeta_2\,(-1+2\,\ln(2))-\zeta_3\right)
\nonumber \\
 + C_F\,T_F \left(
  \frac{115}{9}-8\,\zeta_2\right)+\frac{7}{9} C_F\,T_F\,N_l 
-C_F \beta_0\,\ln(r_Q).
\eea
The term $G_{1,Q}^{(B)}$ is obtained by summing 
over all weak isospin quark doublets in the second order
triangle diagram
contributions to the $Z Q Q$ vertex \cite{Bernreuther:2004ih}.
For definiteness we consider here only the  static
 $t$ and $b$ quark form factors. In determining $G_{1,Q}^{(B)}$
we neglect in both cases the mass of the $b$ quark with respect to
that of the $t$ quark in the respective triangle diagram.
Then we get:
\bea
G_{1,t}^{(B)}(0) = \left(
  \frac{\alpha_s}{2\pi} \right)^2 C_F T_F \left[
  \mathcal{G}_{1,m_t,m_t}(0) 
-  \mathcal{G}_{1,0,m_t}(0)\right] \, , \nonumber \\
G_{1,b}^{(B)}(0) =  \left(
  \frac{\alpha_s}{2\pi} \right)^2 C_F T_F \left[\mathcal{G}_{1,m_b,m_b}(0)
-\mathcal{G}_{1,m_t,0}(0)\right] \, ,\label{g1_s0_bottom}
\eea
where
\bea
\mathcal{G}_{1,m_Q,m_Q}\left(0 \right) & = &
-\frac{19}{3}+\frac{16}{3}\,\zeta_2-3\,\ln(r_Q),
 \nonumber \\
\mathcal{G}_{1,0,m_Q}\left(0 \right) & = &
-7-3\,\ln(r_Q), \nonumber \\
\mathcal{G}_{1,m_Q,0}\left(0 \right) & = &
\frac{3}{2}-3\,\ln(r_Q).
\eea

The induced pseudoscalar form factor $G_{2,Q}(0)$   can also
be computed to order $\alpha_s^2$ from the results of 
\cite{Bernreuther:2004ih}.
However, it is absent in Eq. (\ref{decomp}) for on-shell photons or
$Z$ bosons, 
and irrelevant for the couplings of
heavy quarks to leptons and light quarks via $V$ boson exchange.
Thus it seems not to be
experimentally accessible in the forseeable future. Therefore we shall
not present it here.

\begin{ruledtabular}
\begin{table}[t]
\begin{tabular}{cccc}
 & $t$ quark  &\multicolumn{2}{c}{$b$ quark}\\ \hline
$m_Q$ & 175 GeV & \multicolumn{2}{c}{5 GeV}\\
$Q_Q$ & 2/3 & \multicolumn{2}{c}{-1/3}\\
$T^3_Q$ & 1/2 & \multicolumn{2}{c}{-1/2}\\
\hline
$\mu$ & $m_t$ & $m_b$ & $m_Z$ \\
$\alpha_s(\mu)$ & 0.1080 & 0.2145 & 0.1187\\
\end{tabular}
\caption{Input values used in computing the static top and bottom form
factors. The mass of the $Z$ boson and the weak mixing angle
are taken to be
 $m_Z = 91.187$ GeV and $\sin^2\theta_W = 0.231$
 \cite{Eidelman:2004wy}. $\alpha_s(m_Z)$ is  the ${\overline{\rm MS}}$
QCD coupling 
for  $N_f=5$ flavors, taken from \cite{Eidelman:2004wy}, while $\alpha_s(m_t)$
is the coupling for $N_f=6$ flavors, obtained  from 
$\alpha_s(m_Z)$ by renormalization-group evolution.
\label{tab_werte}}
\end{table}
\end{ruledtabular}
\begin{ruledtabular}
\begin{table}[t]
\begin{tabular}{cccc}
& $t$ $(\mu=m_t)$ & $b$ $(\mu=m_b)$ & $b$
$(\mu=m_Z)$\\ \hline
$(g-2)^{\gamma,(1l)}_Q/2$ & $1.53\cdot 10^{-2}$ & $-1.52\cdot10^{-2}$ & $-8.4\cdot10^{-3}$ \\
 $(g-2)^{\gamma,(2l)}_Q/2$ & $4.7\cdot 10^{-3}$  & $-1.00\cdot10^{-2}$
 & $-6.6\cdot10^{-3}$ \\ \hline
 $(g-2)^{\gamma}_Q/2$      & $2.00\cdot 10^{-2}$ & $-2.52\cdot10^{-2}$
 & $-1.50\cdot10^{-2}$ \\ \hline
 $(g-2)^{Z,(1l)}_Q/2$      & $5.2\cdot 10^{-3}$  & $-1.87\cdot10^{-2}$ & $-1.03\cdot10^{-2}$ \\
 $(g-2)^{Z,(2l)}_Q/2$      & $1.6\cdot 10^{-3}$  & $-1.24\cdot10^{-2}$ & $-8.1\cdot10^{-3}$ \\ \hline
 $(g-2)^{Z}_Q/2$           & $6.8\cdot 10^{-3}$  & $-3.11\cdot10^{-2}$ & $-1.85\cdot10^{-2}$ \\
\end{tabular}
\caption{
One- and two-loop QCD contributions, and their sums, 
to the anomalous magnetic and weak magnetic moments of the top and bottom
quark,
for different values of the renormalization scale $\mu$.
 \label{tab_anomag}}
\end{table}
\end{ruledtabular}
\begin{ruledtabular}
\begin{table}[t]
\begin{tabular}{cccc}
 & $t$ $(\mu=m_t)$ & $b$ $(\mu=m_b)$ & $b$
$(\mu=m_Z)$\\ \hline
 $G_{1,Q}^{(A,1l)}$ & $-2.29\cdot10^{-2}$ & $-4.55\cdot10^{-2}$&$-2.52\cdot10^{-2}$ \\
 $G_{1,Q}^{(A,2l)}$ & $-1.81\cdot10^{-3}$ & $-7.74\cdot10^{-3}$& $-1.30\cdot10^{-2}$\\
 $G_{1,Q}^{(B)}$ & $1.86\cdot10^{-3}$ &$-1.58\cdot10^{-2}$ & $-4.85\cdot10^{-3}$\\ \hline
 $G^Z_{1,Q}/a_Q$ & $1 - 2.29\cdot10^{-2}$
&$1 -  6.90\cdot10^{-2}$ & $1  -  4.31 \cdot10^{-2}$
\end{tabular}
\caption{One- and two-loop QCD contributions
to the static form factor $G^Z_{1,Q}$ as defined in Eq. (\ref{axchar})
for top and
bottom quarks.  \label{tab_schwlad}}
\end{table}  
\end{ruledtabular}
Let us now consider the  static  magnetic and weak magnetic form
factor of a quark $Q$. We consider
\be
\left(\frac{g-2}{2}\right)^{\gamma,Z}_Q \equiv F_{2,Q}^{\gamma,Z}
\left(0 \right) =
{v}_Q^{\gamma,Z}\,F_{2,Q}\left(0 \right) \, ,
\ee
which correspond to the anomalous magnetic (MDM) and weak magnetic
(WMDM) moments of $Q$.
(Notice that in the literature the WMDM is often
associated with  $F_{2,Q}^Z(s=m_Z^2)$.) 
We determine the numerical values of these
moments for $t$ and $b$ quarks with the above formulae and 
the input values given in Table 1.
 In the case of $(g-2)^{\gamma,Z}_t/2$ we work in
$N_f=6$ flavor QCD with all quarks but the top quark taken to
be massless, while $(g-2)^{\gamma,Z}_b/2$ is computed in  the effective
$N_f=5$ flavor theory with $m_i=0$ ($i=u,d,s,c$) and $m_b\neq 0$.
The results are given in Table 2. 
The two-loop QCD
contributions to  $(g-2)^{\gamma,Z}_t/2$ and $(g-2)^{\gamma,Z}_b/2$ 
are sizeable: they 
are about 30 and 70 percent, respectively, of the order
$\alpha_s$ moments. In the case of the $b$ quark we have evaluated the
static magnetic form factors for two different values of the
renormalization scale $\mu$, which would apply to different physical
situations, for instance, the leptoproduction of $b \bar b$ quark
pairs above threshold and at the $Z$ resonance. The dependence on
$\mu$, which is mainly due to the dependence of $\alpha_s$ on this
scale, indicates also the size of the unknown higher-order QCD
corrections. As to the SM electroweak one-loop contributions to the
(W)MDMs of $b$ and $t$ quarks  \cite{Bernabeu}: for the
$b$ quark they are significantly smaller than the QCD-induced moments
given in Table 2. The weak contributions to the WMDM and, if the SM
Higgs boson is light, to the MDM of the $t$ quark are of the order
of a few times $10^{-3}$.

Next we consider the  weak  axial vector charge  $\mathbf{a}_Q^Z$ of a heavy
quark $Q$ defined by
\be
\mathbf{a}_Q^Z \equiv G^Z_{1,Q}\left(0 \right),
\label{axchar}
\ee
where, to second order in the QCD coupling,
$G^Z_{1,Q}$ receives the type A and B contributions given above.
Numerical evaluation of these
formulae  for $t$ and $b$ quarks gives  the values of Table 3.
In the case of the $t$ quark the two-loop type A and B contributions
almost cancel, while for the $b$ quark the
second order corrections  are again sizeable.

What do these numbers tell us? For the $b$ quark
an upper bound  on its magnetic moment 
 was derived in \cite{Escribano:1993xr} from an analysis of LEP1
 data, which, in our
 convention, reads $\delta (g-2)^{\gamma}_b/2 < 1.5 \times
 10^{-2}$ (68 \% C.L.). Comparing it with Table 2 we see that the QCD-induced
 contributions to the $b$ quark magnetic moment saturate this bound, which
 implies that there is limited room for new physics contributions to
this quantity.  At a future linear collider \cite{tesla,Abe:2001nq},
when operated at the $Z$ resonance, the sensitivity to this variable
 could be improved
substantially, either by global fits or by analyzing appropriate
angular distributions in $b {\bar b}$ and $b {\bar b} \gamma$ events. 

As to the static form factors of the  top quark, no such tight
constraints exist so far on possible contributions from new
interactions (see \cite{eboli,Larios:1999au} and \cite{Baur1} for a review). 
As emphasized above these quantities are particularly
sensitive to the dynamics of electroweak symmetry breaking. For
instance, in various models with a strongly coupled symmetry breaking
sector  one may expect contributions from this sector to the static $t$ quark 
form factors at the  5 - 10 \% level
\cite{Chivukula:1995gu,Murayama:1996ec,Hill:2002ap}. The QCD-induced
anomalous magnetic moment and the QCD corrections to the axial charge
of the top quark are of the same order of magnitude. Future colliders
have the potential to reach this level of sensitivity. At the LHC the
$tt\gamma$ and $ttZ$ couplings can be separately measured in
associated $tt\gamma$ and  $ttZ$ production, respectively. A detailed
study of these reactions showed \cite{Baur1,Baur2} that at the LHC a
statistical sensitivity of about 5 \% to the photonic couplings
$F_{1,2,t}^{\gamma}$, $G_{1,t}^{\gamma}$  may eventually be reached,
while the sensitivity to the analogous couplings to the $Z$ boson is 
significantly lower. At a future high luminosity linear $e^+e^-$
collider with polarized beams~\cite{Moortgat-Pick:2005cw}
 both the $Z$ and the photonic couplings
of the top quark will be measurable  with a precision of a few percent
\cite{tesla,Abe:2001nq}. Thus it is mandatory to determine these
quantities within the SM as precisely as possible.

In conclusion, we have determined the static form factors of $b$ and $t$
quarks, notably their anomalous magnetic moments and axial charges, to
second order in the QCD coupling $\alpha_s$. Precise knowledge of
these effective couplings to photons and $Z$ bosons within the
standard model is mandatory in order to assess the margin of
detectablity of new physics contributions and to interpret correctly
(future) measurements.  For the $b$ quark we have found
that the QCD contributions to its anomalous magnetic moment saturate
the existing  experimental upper bound, which implies that there is 
not much room for new physics effects on this quantity.
The QCD corrections to the static form factors of the top quark are of
the same order of magnitude as the precision with
which these couplings may eventually be measured at future colliders and must
therefore be taken into account in searches for anomalous coupling effects.\\

{\bf Acknowledgement:} We thank U.~Baur, A.~Juste and F.~Petriello for
useful discussions.
This work was supported 
 by Deutsche Forschungsgemeinschaft (DFG), 
SFB/TR9, by DFG-Graduiertenkolleg RWTH Aachen, by
HPRN-CT2002-00311 (EURIDICE),
by the Swiss National Science Foundation (SNF) under contract 200021-101874,
and by the USA DoE under the grant DE-FG03-91ER40662, Task J.


\end{document}